# Physicalism Versus Quantum Mechanics

Henry P. Stapp

*Theoretical Physics Group*
*Lawrence Berkeley National Laboratory*
*University of California*
*Berkeley, California 94720*

## Abstract

In the context of theories of the connection between mind and brain, physicalism is the demand that all is basically purely physical. But the conception of "physical" embodied in this demand is characterized essentially by the properties of the physical that hold in classical physical theories. Certain of those properties contradict the character of the physical in quantum mechanics, which provides a better, more comprehensive, and more fundamental account of phenomena. It is argued that the difficulties that have plagued physicalists for half a century, and that continue to do so, dissolve when the classical idea of the physical is replaced by its quantum successor. The argument is concretized in way that makes it accessible to non-physicists by exploiting the recent evidence connecting our conscious experiences to macroscopic measurable synchronous oscillations occurring in well-separated parts of the brain. A specific new model of the mind-brain connection that is fundamentally quantum mechanical but that ties conscious experiences to these macroscopic synchronous oscillations is used to illustrate the essential disparities between the classical and quantum notions of the physical, and in particular to demonstrate the failure in the quantum world of the principle of the causal closure of the physical, a failure that goes beyond what is entailed by the randomness in the outcomes of observations, and that accommodates the efficacy in the brain of conscious intention.

___
**This work was supported by the Director, Office of Science, Office of High Energy and Nuclear Physics, of the U.S. Department of Energy under contract DE-AC02-05CH11231**

# 1. Introduction.

The widely held philosophical position called "physicalism" has been described and defended in a recent book by Jaegwon Kim[1]. The physicalist position claims that the world is basically purely physical. However, "physical" is interpreted in a way predicated, in effect, upon certain properties of classical physics that are contradicted by the precepts of orthodox quantum physics. Kim's arguments reveal two horns of a dilemma that the physicalist is forced to face as a consequence of accepting this classical notion of "physical". Kim admits that neither of the two options, "epiphenomenalism" or "reduction", is very palatable, but he finds a compromise that he deems acceptable.

The central aim of the present paper is to show that the physicalist's dilemma dissolves when one shifts from the classical notion of the physical to the quantum mechanical notion. Understanding this shift involves distinguishing the classical notion of the mind-brain connection from its quantum successor.

To make clear the essential features of the quantum mechanical conception of the mind-brain connection, I shall describe here a model that is a specific realization of a theory I have described in more general terms before[2-5]. Being specific reduces generality, but having a concrete model can be helpful in revealing the general lay of the land. Also, the specific features added here resolve in a natural way the puzzle of how our descriptions of our observations can be couched in the language of classical physics when our brains are operating, fundamentally, in accordance with the principles of quantum theory. The specific model also shows how the thoroughly quantum mechanical (quantum Zeno) effect, which underlies the power of a person's conscious thoughts to influence in useful ways the physically described processes occurring in that person's brain, is not appreciably disrupted either by "environmental decoherence" effects or by thermal effects arising from the "hotness" of the brain.

In order to communicate to the broad spectrum of scientists and philosophers interested in the connection between mind and brain, I will review in the following section the historical and conceptual background of the needed quantum mechanical ideas, and then describe an approach to the mind-body problem that is based fundamentally on quantum theory, but that adds several specific new ideas about the form of the mind-brain connection.

## 2. Quantum Mechanics and Physicalism.

Rather than just plunging ahead and using the concepts and equations of quantum mechanics, and thereby making this work unintelligible to many people that I want to reach, I am going to provide first an historical and conceptual review of the extremely profound changes in the philosophical and technical foundations that were wrought by the transition from classical physics to quantum physics. One key technical change was the shift from the *numbers* used in classical mechanics to describe properties of physical systems to the associated *operators* or *matrices* used to describe related actions. This technical shift emerged, unsought, from a seismic conceptual shift. Following the path blazed by Einstein's success in creating special relativity, Heisenberg changed course. Faced with a quarter century of failures to construct a successful atomic theory based upon the notion of some presumed-to-exist space-time structure of the atom, Heisenberg attempted to build a theory based upon our observations and measurements, rather than upon conjectured microscopic space-time structures that could be *postulated* to exist, but that were never directly observed or measured. This shift in orientation led to grave issues concerning exactly what constituted an "observation" or "measurement". Those issues were resolved by shifting from an ontological perspective---which tries to describe what really exists objectively "out there"--- to a practical or pragmatic perspective, which regards a physical theory as a useful collective conceptual human endeavor that aims to provide us with reliable expectations about our future experiences, for each of the alternative possible courses of action between which we are (seemingly) free to choose. As a collective endeavor, and in that sense as an objective theory, quantum mechanics is built on *descriptions* that allow us to communicate to others what we have done and what we have learned. Heisenberg strongly emphasized that this change in perspective converts the quantum mechanics, in a very real sense, into a theory about "our knowledge": the relationships between experiential elements in our streams of consciousness become the core realities of a conceptual construction that aims to allow us to form, on the basis of what we already know, useful expectations about our future experiences, under the various alternative possible conditions between which we seem able to freely choose.

The paradoxical aspect of claiming the "physical state of a system" to be a representation of "our knowledge" is starkly exhibited by "Schroedinger's cat", whose quantum state is, according to this pragmatic approach, not determined until someone looks. Bohr escapes this dilemma by saying that the current quantum principles are insufficient to cover biological matter, but that approach leaves quantum mechanics fundamentally incomplete, and, in particular, inapplicable to the physical processes occurring in our brains.

In an effort to do better, von Neumann[6] showed how to preserve the rules and precepts of quantum mechanics all the way up to the interface with "experience", thereby preserving the general character of quantum mechanics as a theory that aims to provide reliable expectations about future experiences on the basis of present knowledge. Von Neumann's work brings into sharp focus the central problem of interest here, which is the connection between the properties specified in the quantum mechanical description of a person's brain and the experiential realities that populate that person's stream of consciousness. Bohr was undoubtedly right in saying that the orthodox precepts would be insufficient to cover this case. Additional ideas are needed, and the purpose of this article is to provide them.

The switch from classical mechanics to quantum mechanics preserves the idea that a physical system has a physically describable state. But the character of that state is changed drastically. Previously the physical state was conceived to have a well defined meaning independently of any "observation". Now the physically described state has essentially the character of a "potentia" (an "objective tendency") for the occurrence of each one of a continuum of alternative possible "events". Each of these alternative possible events has both an experientially described aspect and also a physically described aspect: each possible "event" is a psycho-physical happening. The experientially described aspect of an event is an element in a person's stream of consciousness, and the physically described aspect is a *reduction* of the set of objective tendencies represented by the prior state of that person's body-brain to the *part* of that prior state that is compatible with the increased knowledge supplied by the new element in that person's stream of consciousness. Thus the changing psychologically described state of that person's knowledge is correlated to the changing physically described state of the person's body-brain, and the changing physically described state entails, via the fundamental quantum probability formula, a changing set of weighted possibilities for future psycho-physical events.

The practical usefulness of quantum theory flows from this lawful connection between a person's increasing knowledge and the changing physical state of his body-brain. The latter is linked to the surrounding physical world by the dynamical laws of quantum physics. This linkage allows a person to "observe" the world about him by means of the lawful relationship between the events in his stream of conscious experiences and the changing state of his body-brain.

It is worth noting that the physically described aspect of the theory has lost its character of being a "substance", both in the philosophical sense that it is no longer *self-sufficient,* being intrinsically and dynamically linked to the mental, and also in the colloquial sense of no longer being *material*. It is *stripped of materiality* by its character of being merely a potentiality or possibility for a future event. This shift in its basic character renders the physical aspect somewhat idea-like, even though it is conceived to represent objectively real tendencies.

The key "utility" property of the theory---namely the property of being useful---makes no sense, of course, unless we have, in some sense, some freedom to choose. An examination of the structure of quantum mechanics reveals that the theory has both a logical place for, and a logical need for, choices that are made in practice by the human actor/observers, but that are *not determined by the quantum physical state of the entire world, or by any part of it*. Bohr[7] calls this choice "the free choice of experimental arrangement for which the quantum mechanical formalism offers the appropriate latitude." (Bohr, p.73). This "free" choice plays a fundamental role in von Neumann's rigorous formulation of quantum mechanics, and he gives the physical aspect of this probing action the name "process 1" (von Neumann, p. 351, 418, 421). This process 1 action is not determined, even statistically, by the physically described aspects of the theory.

The fact that this choice made by the human observer/agent is not determined by the physical state of the universe means that *the principle of the causal closure of the physical domain is not maintained in contemporary basic physical theory*. It means also that Kim's formulation of *mind-body supervenience is not entailed by contemporary physical theory*. That formulation asserts that "what happens in our mental life is wholly dependent on, and determined by, what happens with our bodily processes." (Kim, p. 14) Kim indicates that supervenience is a common element of all *physicalist theories.* But since supervenience is not required by basic (i.e., quantum) physics, the easy way out of the difficulties that have been plaguing physicalists for half a century, and that continue to do so, is simply to recognize that the precepts of classical physics, which are the scientific source of the notions of the causal closure of the physical, and of supervenience, do not hold in real brains, whose activities are influenced heavily by quantum processes.

Before turning to the details of the quantum mechanical treatment of the relationship between mind and brain I shall make a few comments on Kim's attempted resolution of the difficulties confronting the classical physicalist approach. The essential problem is the mind-body problem. Kim divides this problem into two parts, the problem of mental causation and the problem of consciousness. The problem of mental causation is: "How can the mind exert its causal powers in a world that is fundamentally physical?" The problem of consciousness is: "How can a thing such as consciousness exist in a physical world, a world consisting ultimately of nothing but bits of matter distributed over space-time in accordance with the laws of physics."

From a modern physics perspective the way to resolve these problems is immediately obvious: Simply recognize that the assumption that the laws of physics pertain to "bits of matter distributed over space-time in accordance with the laws of physics" is false. Indeed, that idea has, for most of the twentieth century, been asserted by orthodox physicists to be false, along with the assumption that the world is physical in the classical sense. Quantum mechanics builds upon the obvious real existence of our streams of conscious experiences, and provides also, as we shall see, a natural explanation of their causal power to influence physical properties. Thus the difficulties that have beset physicalists for five decades, and have led to incessant controversies and reformulations, stem, according to the perspective achieved by twentieth century physics, directly from the fact that the physicalist assumptions not only do not follow from basic precepts of physics, but instead, directly contradict them. The premises of classical physicalists have been, from the outset, incredibly out of step with the physics of their day.

Kim tries at one point to squash the notion that the difficulties with physicalism can be avoided by accepting some form of dualism. But the dualism that he considers is a Cartesian dualism, populated with mysterious souls. However, quantum mechanics is science! The experientially described realities that occur in quantum theory are the core realities of science. They are the ideas that we are able communicate to others pertaining to what we have done and what we have learned. These descriptions are essentially descriptions of (parts of) the accessible contents of the streams of consciousness of real living observer-agents. Criticizing dualism in the form advanced by Descartes during the seventeenth century instead of in the form employed in contemporary science is an indication that philosophers of mind have isolated themselves in a hermetically sealed world, created by considering only what other philosophers of mind have said, or are saying, with no opening to the breezes that bring word of the highly pertinent revolutionary change that had occurred in basic science decades earlier.

Kim's main argument leads to the conclusion that a physicalist must, for each conscious experience, choose between two options: either that experience is causally powerless, or it must be *defined to be* the causally efficacious brain activity that possesses its causal power. Kim himself admits that neither option is very palatable. The idea that our beliefs, desires, and perceptions, including our pains, have no effects upon our actions is regarded by Kim as unacceptable. Thus he opts for what he claims to be the only alternative available to the rational physicalist, namely that each such efficacious experience must be (defined to be) a causally efficacious brain activity that causes its effects: "If anything is to exercize causal power in the physical domain it must be an element in the physical domain or be reducible to it." (Kim, p, 170-171) "Only physically reducible mental properties can be causally efficacious." (Kim, p. 174)

That a conscious experience can be defined to be a physical activity, described in the mathematical language of physics, is certainly a hard pill to swallow. Fortunately, it is not true in quantum theory, where the physically described state represents merely an "objective tendency' for a psych-physical event to occur. However, the mind-brain identity that Kim describes does have a less-problematic *analog* in quantum theory. Each actual event has two sides: an experience; and a reduction of the prior state of the body-brain to one that incorporates into the physically described world a causal aspect conceptually represented in the intentional aspect of the experience. This is the essential core of the orthodox von Neumann/Heisenberg quantum position. It will be elaborated upon here.

Kim's solution has another apparent defect: different aspects of a person's apparently highly integrated stream of consciousness have fundamentally different statuses, in regard to their connections to that person's brain. Beliefs, desires and percepts are defined to be brain activities, whereas colors and other "qualia" are not brain activities and are not causally efficacious. But how can your desire for a beautiful painting be simply a brain activity, whereas the particular colors that combine to excite this desire are epiphenomenal qualities having no effects on your brain?

The physicalist assumption has apparently led, after 50 years of development, to conclusions that are far from ideal. These conclusions fail to explain either why our conscious experiences should exist at all in a world that is dynamically and logically complete at the physical level of description, or how they can *be* physical properties that do not entail the existence of the experiential "feel" that characterize them. These long-standing difficulties arise directly from accepting the classical conception of the nature and properties of the physically described aspects of our description of the world. They are resolved in a natural way by accepting the quantum mechanical conception of the nature and properties of the aspects of the world that are described in physical terms: i.e., in terms of properties specified by assigning mathematically properties to space-time regions.

In the following sections I shall explain how these difficulties are resolved by accepting the quantum conception of the physical.

3. **Quantum Mechanics: The Rules of the Game.**

3.1 The basic formula.

Quantum mechanics is a superstructure erected upon a basic formula. This formula specifies the probability that a probing action that is describable in everyday language, refined by the concepts of classical physical theory, will produce a *pre-specified* possible experienced outcome that is described in the same kind of terms. First a *preparing* action must be performed. Its outcome is represented by a (quantum) state of the prepared system. Then a probing action is chosen and performed. The elementary probing actions are actions that either produce a pre-specified outcome 'Yes', or that fail to produce that pre-specified outcome.

To achieve generality I shall adopt the density matrix formulation described by von Neumann. In this formulation the physical state of a system is represented by a matrix that is called the density matrix. It is traditionally represented by the symbol $\rho$. A measurement or observation on such a system is effected by means of a probing action, which is represented by a matrix, traditionally designated by the symbol *P*, or by a *P* with a subscript, that satisfies *PP=P*. Such a matrix/operator is called a *projection operator*. The quantum game is like "twenty questions": the observer-agent "freely poses" a question with an observable answer 'Yes' or 'No', This question, and the probing action corresponding to it, are represented in the formalism by some projection operator *P*. Nature then returns an answer 'Yes' or 'No'. The probability that the answer is 'Yes' is given by the basic probability equation of quantum mechanics:

*<P> = Trace P$\rho$/Trace $\rho$.*

In order not to lose non-physicists, but rather to get them into the quantum swing of things, and allow them to play this wonderful game, I shall spell out what this equation means in the simple case in which the matrices involved have just two rows and two columns. In this case each matrix/operator has four elements, which are specified by the four numbers *<1|M|1>, <1|M|2>, <2|M|1>,* and *<2|M|2>*. The index on the left specifies the horizontal row, and the index on the left specifies the vertical column of the matrix in which the matrix element is to be placed. The rule of matrix multiplication says, for any two matrices M and N, and any pair of two-valued indices *i and j,*

*<i|MN|j> = <i|M|k><k|N|j>,*

where one is supposed to sum over the two possible values of the repeated index k. For any M,

*Trace M = <k|M|k>,*

where one is again supposed to sum over the (two in this case) different possible values (1 and 2) of the index *k*.

This case of a system represented by two-by-two matrices is physically very important: it covers the case of the "spin" degree of freedom of an electron. Once one sees how quantum mechanics works in this simplest case, the generalization to all other cases is basically pretty obvious. So in order to keep non-physicists on board I will spend a little time spelling things out in detail for this simple case.

Pauli introduced for this two-by-two case four particular matrices defined by

$<1| \sigma_0 |1> = 1$, $<1| \sigma_0 |1> = 1$, $<1| \sigma_1 |2> = 1$, $<2| \sigma_1 |1> = 1$,
$<1| \sigma_2 |2> = -i$, $<2| \sigma_2 |1> = i$, $<1| \sigma_3 |1> = 1$,
$<2| \sigma_3 |2> = -1$;

with all other elements zero. [ $i$ is the imaginary unit]

They satisfy $\sigma_j \sigma_j = \sigma_0 = I$, for all $j$; $\sigma_1 \sigma_2 = i\sigma_3 = -\sigma_2 \sigma_1$; $\sigma_2 \sigma_3 = i\sigma_1 = -\sigma_3 \sigma_2$; and $\sigma_3 \sigma_1 = i\sigma_2 = -\sigma_1 \sigma_3$. Most calculations can be done using just these products of the Pauli matrices.

For actions that probe the direction of the spin of the electron the projection operator $P = \frac{1}{2}(I + \sigma_3)$ represents the probing action that corresponds to the query "Does the spin of the electron point in the direction of the axis number 3?" It is also the density matrix that represents the spin state of the electron if the answer to that query is 'Yes'. In the higher dimensional cases, if ρ is the density matrix prior to the probing action then the density matrix after a probing action $P$ that produces the answer 'Yes' is PρP (up to a possible positive multiplicative factor that drops out of the probability formula.)  If the feedback is 'No' then ρ is reduced to *P' ρ P'*, with *P'= (1-P).*

Suppose one has prepared the spin state of the electron by performing the probing action corresponding to $P = \frac{1}{2}(I + \sigma_3)$ and has received the answer 'Yes'. This means that the density matrix for the system is now (known to be the state represented by) ρ = $\frac{1}{2}(I + \sigma_3)$. Suppose one now performs the probing action corresponding to the query "Does the spin point in the direction of axis number 1. The corresponding P is $\frac{1}{2}(I + \sigma_1)$. Thus the probability that the answer is 'Yes' is

*Trace* $\frac{1}{2}(I + \sigma_1)$ $\frac{1}{2}(I + \sigma_3)$/*Trace*$\frac{1}{2}(I + \sigma_3) = 1/2$.

This simplest example beautifully epitomizes the general case. It illustrates very accurately how the basic probability formula is used in actual practice.

The basic probability formula and its workings constitute the foundation of the quantum mechanical conception of the connection between the aspects of our scientific understanding of nature described in the language that we use to describe the pertinent perceptual and felt contents of our streams of conscious experiences and the aspects described in the mathematical language of physics.

3.2 Classical Description.

"…we must recognize above all that, even when phenomena transcend the scope of classical physical theories, the account of the experimental arrangement and the recording of observations must be given in plain language, suitably supplemented by technical physical terminology. This is a clear logical demand, since the very word "experiment" refers to a situation where we can tell others what we have done and what we have learned." (Bohr, p. 72)

"…it is imperative to realize that in every account of physical experience one must describe both experimental conditions and observations by the same means of communication as the one used in classical physics." (Bohr, p. 88)

This demand that we *must use* the known-to-be-fundamentally-false concepts of classical physical theories as a fundamental part of quantum mechanics has often been cited as the logical incongruity that lies at the root of the difficulties in arriving at a rationally coherent understanding of quantum mechanics: of an understanding that goes beyond merely understanding how to use it in practice. So I focus next on the problem of reconciling the quantum and classical concepts, within the context of a theory of the mind-brain connection.

3.3 Quasi-Classical States of the Electromagnetic Field

There is one part of quantum theory in which a particularly tight and beautiful connection is maintained between classical mechanics and quantum mechanics. This is the simple harmonic operator (SHO). With a proper choice of units the energy (or Hamiltonian) of the system has the simple quadratic form $E = H = \frac{1}{2}(p^2 + q^2)$, where q and p are the coordinate and momentum *variables* in the classical case, and are the corresponding *operators* in the quantum case. In the classical case the trajectory of the "particle" is a circle in q-p space of radius $r = (2E)^{1/2}$. The angular velocity is constant and independent of E, and in these special units is $\omega = 1$: one radian per unit of time. The lowest-energy classical state is represented by a point at rest at the "origin" q = p = 0.

The lowest-energy quantum state is the state---i.e., projection operator P---corresponding to a Gaussian wave function that in coordinate space is
ψ(q) = C exp(— (½)q$^2$) and in momentum space is
ψ(p) = C exp(— (½ )p$^2$), where C is $2^{1/4}$. If this ground state is shifted in q-p space by a displacement (Q, P) one obtains a state---i.e., a projection operator P---labeled by [Q,P], which has the following important property: if one allows this quantum state to evolve in accordance with the quantum mechanical equations of motion then it will evolve into the set of states labeled by [Q(t), P(t)], where the (center) point (Q(t), P(t)) moves on a circular trajectory that is identical to the one followed by the classical point particle.

If one puts a macroscopic amount of energy E into this quantum state then it becomes "essentially the same as" the corresponding classical state. Thus if the energy E in this one degree of freedom is the energy per degree of freedom at body temperature then the quantum state, instead of being confined to an *exact point* (Q(t), P(t)) lying on a circle of (huge) radius r = $10^{15}$ in q-p space, will be effectively confined, due to the Gaussian fall-off of the wave functions, to a disc of unit radius centered at that point (Q(t), P(t)). Given two such states, [Q,P], and [Q',P'], their overlap, defined by the Trace of the product of these two projection operators, is exp(— (½ )d$^2$), where d is the distance between their center points. On this $10^{15}$ scale the unit size of the quantum state becomes effectively zero. And if the energy of this classical SHO state is large on the thermal scale then its motion, as defined by the time evolution of the projection operator [Q(t), P(t)] = P$_{[P(t),Q(t)]}$, will be virtually independent of the effects of both environmental decoherence, which arises from subtle quantum-phase effects, and thermal noise, for reasons essentially the same as the reasons for the negligibility these effects on the classically describable motion of the pendulum on a grandfather clock.

Notice that the quantum state [Q, P] is completely specified by the corresponding classical state (Q, P): the quantum mechanical spreading around this point is not only very tiny on the classical scale; it is also completely fixed: the width of the Gaussian wave packet associated with our Hamiltonian is fixed, and independent of both the energy and phase of the SHO.

We are interested here in brain dynamics. Everyone admits that at the most basic dynamical level the brain must be treated as a quantum system: the classical laws fail at the atomic level. This dynamics rests upon myriads of microscopic processes, including flows of ions into nerve terminals. These atomic-scale processes must in principle be treated quantum mechanically. But the effect of accepting the quantum description at the microscopic level is to inject quantum uncertainties/indeterminacies at this level. Yet introducing even small uncertainties/indeterminacies at microscopic levels into these nonlinear systems possessing lots of releasable stored chemical energy has a strong tendency---the butterfly effect---to produce very large macroscopic effects later on. Massive parallel processing at various stages may have a tendency to reduce these indeterminacies, but it is pure wishful thinking to believe that these indeterminacies can be completely eliminated in all cases, thereby producing brains that are completely deterministic at the macroscopic level. *Some* of the microscopic quantum indeterminacy *must* at least occasionally make its way up to the macroscopic level.

According to the precepts of orthodox quantum mechanics, these macroscopic quantum uncertainties are resolved by means of process 1 interventions, *whose forms are not specified by the quantum state of the universe, or any part thereof*. What happens in actual practice is determined by conscious choices "for which the quantum mechanical formalism offers the appropriate latitude". No way has yet been discovered by quantum theorists to circumvent this need for some sort of intervention that is not determined by the orthodox physical laws of quantum physics. In particular, environmental decoherence effects certainly do not, by themselves, resolve this problem of reconciling the quantum indeterminacy, which irrepressibly bubbles up from the microscopic levels of brain dynamics, with the essentially classical character of our descriptions of our experiences of "what we have done and what we have learned".

The huge importance of the existence and properties of the quasi-classical quantum states of SHOs is this: If the projection operators P associated with our experiences are projection operators of the kind that instantiate these quasi-classical states then we can rationally reconcile the demand that the dynamics of our brains be fundamentally quantum mechanical with the demand that our descriptions of our experiences of "what we have done and what we have learned" be essentially classical. This arrangement would be a natural upshot of the fact that our experiences correspond to the actualization of strictly quantum states that are both specified by classical states, and that also closely mimic the properties of their classical counterparts, *apart from the fact that they represent only potentialities*, and hence will be subject, just like Schroedinger' macroscopic cat, to the actions of the projection operators associated with our probing actions. This quantum aspect entails that, by virtue of the quantum Zeno effect, which follows from the basic quantum formula that connects our conceptually described observations to physically described quantum jumps, we can understand *dynamically* how our conscious choices can affect our subsequent thoughts and actions: we can rationally explain, by using the basic principles of orthodox contemporary physics, the causal efficacy of our conscious thoughts in the physical world, and thereby dissolve the physicalists' dilemma.

I shall now describe in more detail how this works.

**4. The Mind-Brain Connection.**

The general features of this quantum approach to the mind-brain problem have been described in several prior publications[2-5,8-11]. In this section I will present a specific model based on the general ideas described in those publications.

Mounting empirical evidence[12,13] suggests that our conscious experiences are connected to brain states in which measurable components of the electromagnetic field located in spatially well separated parts of the brain are oscillating with the same frequency, and in phase synchronization. The model being proposed here assumes, accordingly, that the brain correlate of each conscious experience is an EM (electromagnetic) excitation of this kind. More specifically, each process 1 probing action is represented quantum mechanically in terms of a projection operator that is the quasi-classical counterpart of such an oscillating component of a classical EM field.

The central idea of this quantum approach to the mind-brain problem is that each process 1 intervention is the physical aspect of a psycho-physical event whose psychologically described aspect is the conscious experience of intending to do, or choosing to do, some physical or mental action. The physical aspect of the 'Yes' answer to this probing event is the actualization, by means of a quantum reduction event, of a pattern of brain activity called a "template for action". A *template for action* for some action X is a pattern of physical (brain) activity which if held in place for a sufficiently long time will tend to cause the action X to occur. The psycho-physical linkage between the conscious intent and the linked template for action is supposed to be established by trial and error learning.

A prerequisite for trial and error learning of this kind is that mental effort be causally efficacious in the physically described world. Only if conscious choices and efforts have consequences in the physically described world can an appropriate correlation connecting the two be mechanically established by trial and error learning. With no such connection the conscious intention could become completely opposed to the correlated physical action, with no way to activate a corrective physical measure.

The feature of quantum mechanics that allows a person's conscious choices to influence that person's physically described brain process in the needed way is the so-called "Quantum Zeno Effect". This quantum effect entails that if a sequence of very similar process 1 probing actions occur in sufficiently rapid succession then the affected component of the physical state will be forced, with high probability, to be, at the particular sequence of times $t_i$ at which the probing actions are made, exactly the sequence of states specified by the sequence of projection operators $P_h(t_i)$ that specify the 'Yes' outcomes of the sequence of process 1 actions. That is, the affected component of the brain state---for example some template for action---will be forced, with high probability, *to evolve in lock step with a sequence of 'Yes' outcomes of a sequence of "freely chosen" process 1 actions, where "freely chosen" means that these process 1 actions are not determined, via any known law, by the physically described state of the universe!* This coercion of a physically described aspect of a brain process to evolve in lock step with the 'Yes' answers to a sequence of process 1 probing actions that are free of any known physically described coercion, but that seem to us to be freely chosen by our mental processes, is what will presently be demonstrated. It allows physically un-coerced conscious choices to affect a physically described process that will, by virtue of the basic probability formula, have experiential consequences.

The repetition rate (attention density!) in the sequence of process 1 actions is assumed to be controlled by conscious effort. In particular, in the model being described here, where the projection operators $P(t_i)$ are projection operators $[Q(t_i), P(t_i)]$ that are quasi-classical states of SHOs, the size of the intervals ($t_{i+1} - t_i$) --- being a feature of the sequence of "freely chosen" process 1 probing action --- is taken to be under the immediate control of the psychological aspect of the probing action.

I describe the quantum properties of the EM field in the formulation of relativistic quantum field theory developed by Tomonaga[14] and Schwinger[15], which generalizes the idea of the Schroedinger equation to the case of the electromagnetic field. One can imagine space to be cut up into very tiny regions, in each of which the values of the six numbers that define the electric and magnetic fields in that region are defined. In case the field in that region is executing simple harmonic oscillations we can imagine that each of the six values is moving in a potential well that produces the motion of a SHO. If the process 1 action is specified by a 'Yes' state that is a coordinated synchronous oscillation of the EM field in many regions, {$R_1$, $R_2$, $R_3$, …} then this state, if represented quantum mechanically, consists of some quasi-classical state [$Q_1$, $P_1$] in $R_1$, *and* some quasi-classical state [$Q_2$, $P_2$] in $R_2$, *and* some quasi-classical [$Q_3$, $P_3$] in $R_3$, etc.. The state P of this combination is the *product* of these [$Q_i$, $P_i$]s, each of which acts in its own SHO space, and acts like the unit operator (i.e., unity or 'one') in all the other spaces. This product of $P_n$ s, all evaluated at time $t_i$, is the $P_h(t_i)$ that is the brain aspect of the 'Yes' answer to the process 1 query that occurs at time $t_i$. The *quantum* frequency of the state represented by this $P_h(t_i)$ is the sum of the *quantum* frequencies of the individual regions, and is the total number of quanta in the full set of SHOs. However, the period of the periodic motion of the classical EM field remains $2\pi$, in the chosen units, independently of how many regions are involved, or how highly excited the states of the SHOs in the various regions become. This smaller frequency is the only one that the classical state knows about: it is the frequency that characterizes the features of brain dynamics observed in EEG and MEG measurements.

The sequence of $P_h(t_i)$s that is honed into observer/agent's structure by trial and error learning is a sequence of $P_h(t_i)$s that occurs when the SHO template for action is held in place by effort. Learning is achieved by effort, which increases attention density, and holds the template for action in place. Thus if $H_0$ is the Hamiltonian that maintains this SHO motion then for the honed sequence

$$P_h(t_{i+1}) = \exp(-iH_0(t_{i+1} - t_i)) P_h(t_i) \exp(iH_0(t_{i+1} - t_i)).$$

But in the new situation there may be disturbing physical influences that tend to cause a deviation from the learned SHO motion. Suppose that on the time scale of $(t_{i+1} - t_i)$ the disturbance is small, so that the perturbed evolution starting from $P_h(t_i)$ can be expressed in the form

$$P(t_{i+1}) = \exp(-iH_i(t_{i+1} - t_i)) \exp(-iH_0(t_{i+1} - t_i)) P_h(t_i)$$
$$\exp(iH_0(t_{i+1} - t_i)) \exp(iH_i(t_{i+1} - t_i))$$

$$= \exp(-iH_i(t_{i+1} - t_i))) P_h(t_{i+1}) \exp(iH_i(t_{i+1} - t_i))$$

where $H_i$ is bounded.

According to the basic probability formula, the probability that this state $P(t_{i+1})$ will be found, if measured/observed, to be in the state $P_h(t_{i+1})$ at time $t_{i+1}$ is (using Trace $P_h(t_i) = 1$)

Trace $P_h(t_{i+1}) \exp(-iH_i(t_{i+1} - t_i))) P_h(t_{i+1}) \exp(iH_i(t_{i+1} - t_i))$.

Inserting the leading and first order terms $[1 \pm iH_i(t_{i+1} - t_i)]$ in the power series expansion of $\exp(\pm iH_i(t_{i+1} - t_i))$ and using PP= P, and the fact that Trace AB = Trace BA, for all A and B, one finds that the term linear in $(t_{i+1} - t_i)$ vanishes identically.

The vanishing of the term linear in $(t_{i+1} - t_i)$ is the basis of the quantum Zeno effect. If one considers some finite time interval and divides it into small intervals $(t_{i+1} - t_i)$ and looks at a product of factors $(1 + c(t_{i+1} - t_i)^n)$, then if n is bigger than one the product will tend to unity (one) as the size of the intervals $(t_{i+1} - t_i)$ tend to zero. But this means that the basic probability formula of quantum mechanics requires that, as the step sizes $(t_{i+1} - t_i)$ tend to zero, the evolving state of the system being probed by the sequence of probing action will have a probability that tends to one (unity) *to evolve in lock step with the set of 'Yes' answer to the sequence of probing actions*, provided the initial answer was 'Yes'. But the forms of the projection operators $P_h(t_i)$ and the timings of the probing actions are not determined by the laws of orthodox quantum theory: they are "freely chosen". Hence orthodox quantum theory accommodates in natural way the capacity of a person's conscious intentional choices to influence the processes occurring in his or her physically described brain, and to influence them in a way that will tend to produce intended consequences.

The point of his derivation is that it is expressed in terms of brain states that are macroscopic, and that correspond to classically describable states of the electromagnetic field measured by EEG and MEG procedures. Even though these states contain huge amounts of energy, nevertheless, if we accept the principle that the underlying brain dynamics must in principle be treated quantum mechanically, and, accordingly, replace these classical states by their quasi-classical counterparts, which represent potentialities that are related to experience only via the basic equation, then the principles of orthodox von Neumann quantum mechanics provide a rationally coherent way of understanding the mind-brain connection in a way that escapes the horns of the physicalists' dilemma: it gives each person's intentional conscious choices the power to causally effect the course of events in his or her quantum mechanically described brain, and to influence it in a way that serves these intentions.

## Acknowledgement

I thank Ed Kelly for many useful suggestions pertaining to the form of this paper.

## References

1. J. Kim, *Physicalism, Or Something Near Enough*, (Princeton University Press, Princeton, NJ, 2005)

2. H. P. Stapp, *Mind, Matter, and Quantum Mechanics*, (Springer, Berlin & New York, 2004) [Second Edition]

3. H. P. Stapp, *Mindful Universe: Quantum Mechanics and the Participating Observer*, (Springer, Berlin & New York, 2007).

4. H.P. Stapp, Quantum Interactive Dualism: An Alternative to Materialism, *J. Consc. Studies*. 12 no.11 43-58 (2005). [http://www-physics.lbl.gov/~stapp/stappfiles.html]

5. J. M. Schwartz, H.P. Stapp, & M. Beauregard, Quantum theory in neuroscience and psychology: A neurophysical model of the mind/brain interaction. *Phil. Trans. Royal Soc.* B 360 (1458) 1306 (2005).

6. J. von Neumann, *Mathematical Foundations of Quantum Mechanics*. (Princeton University Press, Princeton, NJ, 1955) [Translated from the 1932 German original by R. T. Beyer]

7. Bohr, N, *Atomic Physics and Human Knowledge*, (Wiley, New York, 1958)

8. H. P. Stapp in *Physics and Whitehead: Quantum. Process, and Experience*, eds. T. Eastman & H. Keeton, (SUNY press, Albany NY, 2004)

9. H.P. Stapp, Light as Foundation of Being, in *Quantum Implications: Essays in Honor of David Bohm*, (Routledge and Kegan Paul, London & New York, 1987).

10. H.P. Stapp, On the unification of quantum theory and classical physics, in *Symposium on the Foundations of Modern Physics: 50 years of the Einstein-Podolsky-Rosen Gedankenexperiment,* eds. P. Lahti and P. Mittelsteadt, (World Scientific, Singapore, 1985).

11. J.R. Klauder & E.C.G. Sudarshan, *Fundamentals of Quantum Optics,* (W.A. Benjamin, New York, 1968). [The quasi-classical states used in the present paper are the projection operators corresponding to the the "coherent states" described in this reference.]

12. J. Fell, G. Fernandez, P. Klaver, C. Elger, & P.Fries, Is synchronized neuronal gamms activity relevant for selective attention? *Brain Res Rev* **42** (2003) 265-272.

13. A. Engel, P Fries, & W. Singer, Dynamic predictions: Oscillations and synchrony in top-down processing, *Nat Rev Neurosci* **2** (2001) 704-716.

14. Tomonaga, S. On a relativistically invariant formulation of the quantum theory of wave fields. *Progress in Theoretical Physics*, **1**, (1946) 27-42.

*15*. Schwinger, J. Theory of Quantized fields I, *Physical Review*, **82**, (1951) 914-927.